\documentclass{camera}
\usepackage{graphicx}  
\newcommand{\be}{\begin{equation}}
\newcommand{\ee}{\end{equation}}
\newcommand{\bea}{\begin{eqnarray}}
\newcommand{\eea}{\end{eqnarray}}

\begin{document}

%
\title{Bottomonium melting at temperature well above Tc}

%
\author{Pedro Bicudo, Jo\~ao Seixas \and Marco Cardoso}

%
\organization{CFTP, Dep. F\'{\i}sica,
Instituto Superior T\'ecnico,
\\
Av. Rovisco Pais, 1049-001 Lisboa, Portugal}

\maketitle

\begin{abstract}
We fit the lattice QCD data of Kaczmarek et al for the free energy F1 and internal energy U1 with a class of
Coulomb, constant and linear potentials, both pure and screened, matching the large
distance to the short distance parts of the lattice QCD finite temperature energies. We also include the hyperfine potential in F1 and U1. We
detail the bottomonium (and charmonium) binding and the melting temperatures, both for the groundstate and for the excited states, relevant for Hard Probes in Heavy Ion Collisions at LHC,
where temperatures well above Tc will be reached, much higher than the temperatures
reached in previous Heavy Ion Collisions.
\end{abstract}

%
Bottomonium and charmonium are good prototypes to study finite T quark-antiquark potentials 
\cite{Matsui:1986dk}
since
\be
	m_b, m_c >> {\Lambda}_{QCD} , Tc
\ee
allow us to neglect
 spontaneous chiral symmetry breaking,
 relativistic effects,  coupled channels,
and temperature in the quark propagators.
In this short contribution we simply aim, in the spirit of \cite{Wong:2004zr}, to solve the Schr\"odinger equation with static lattice QCD potentials.

\begin{figure}[t]
\resizebox{0.5\textwidth}{!}{\includegraphics[width=0.5\columnwidth]{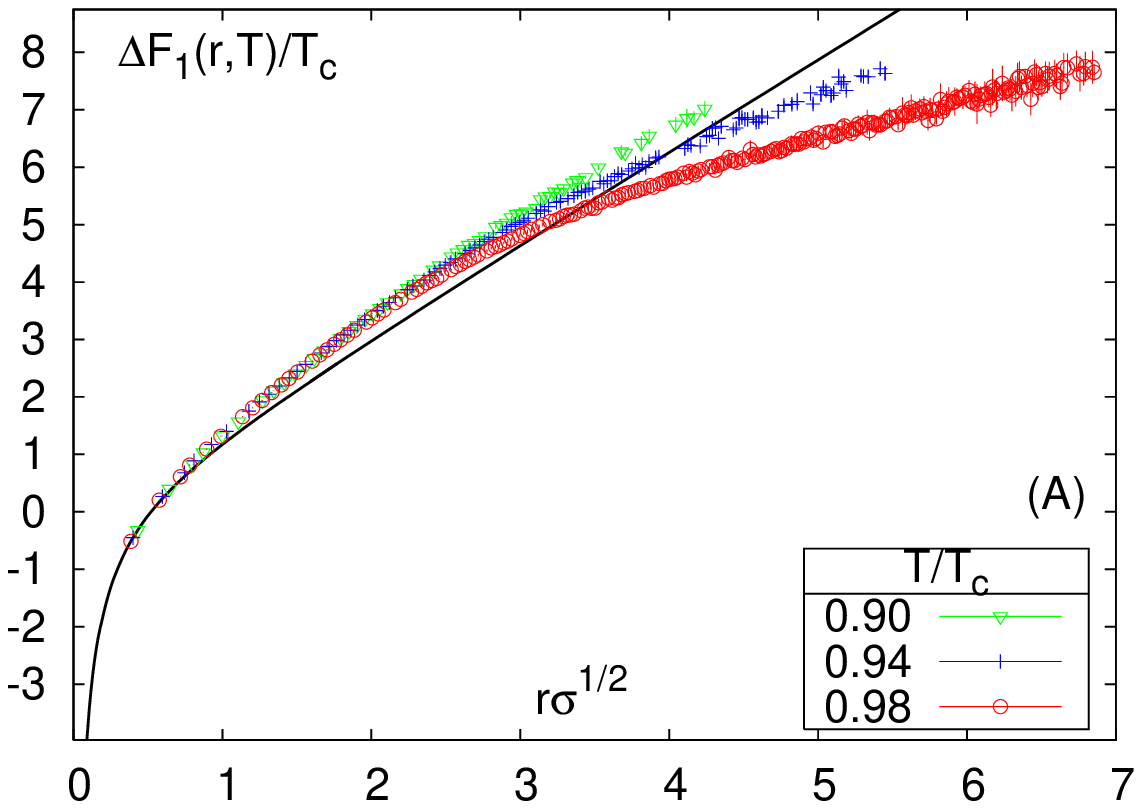}}
\resizebox{0.5\textwidth}{!}{\includegraphics[width=0.5\columnwidth]{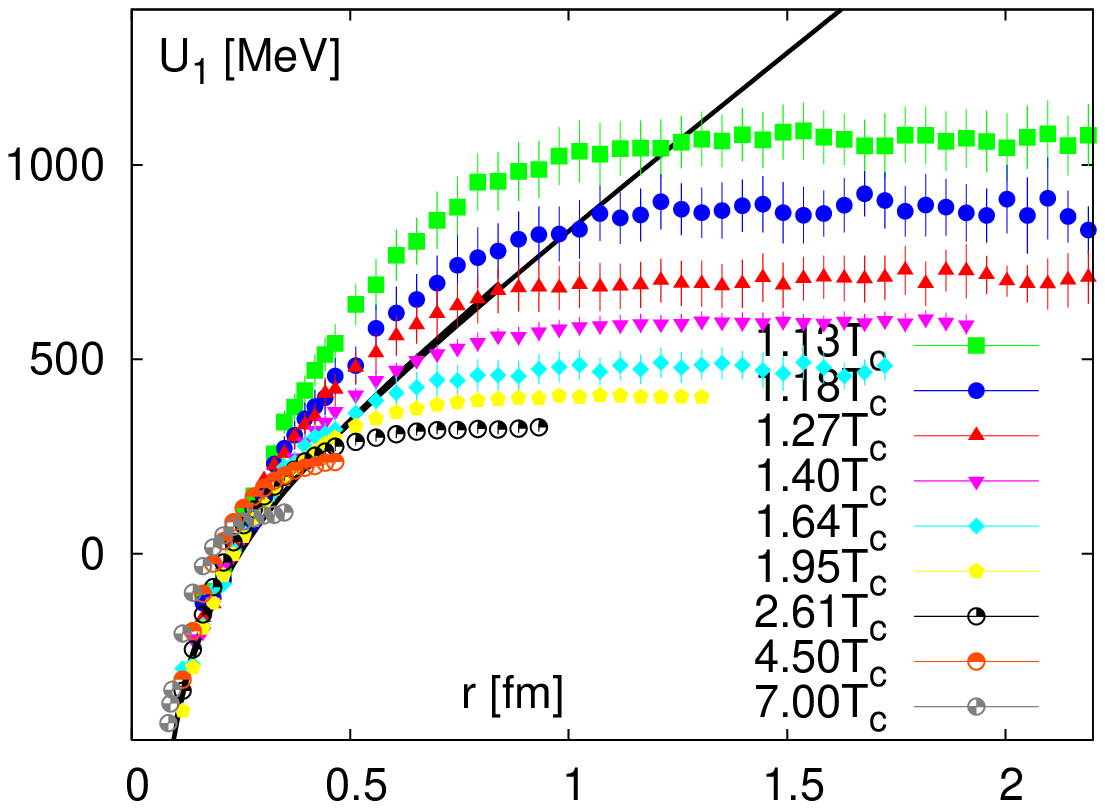}}
\\
\vspace{-1.5 cm}
\\
\resizebox{0.6\textwidth}{!}{\includegraphics[width=0.6\columnwidth]{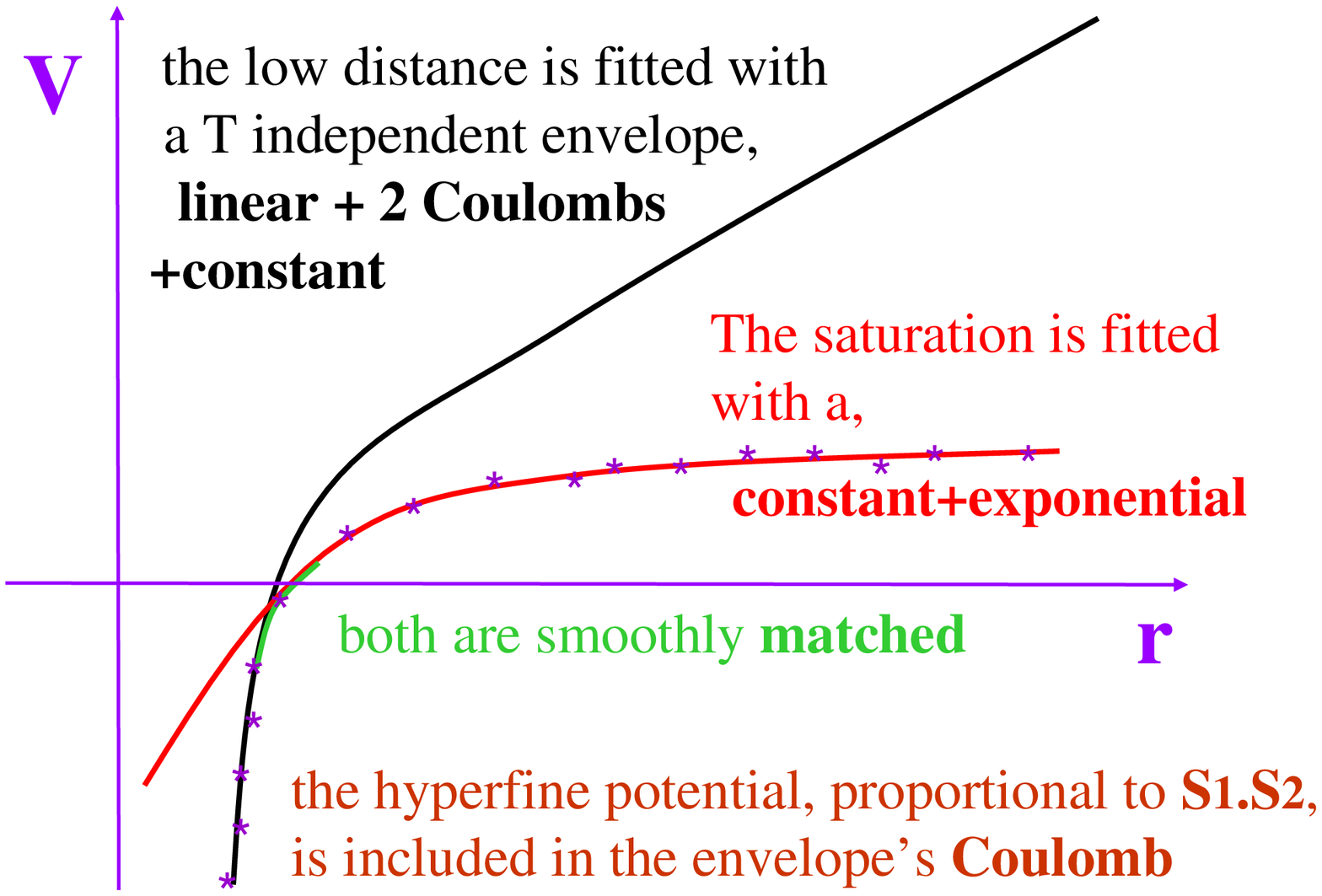}}
\hspace{-1cm}
\resizebox{0.45\textwidth}{!}{\includegraphics[width=0.4\columnwidth]{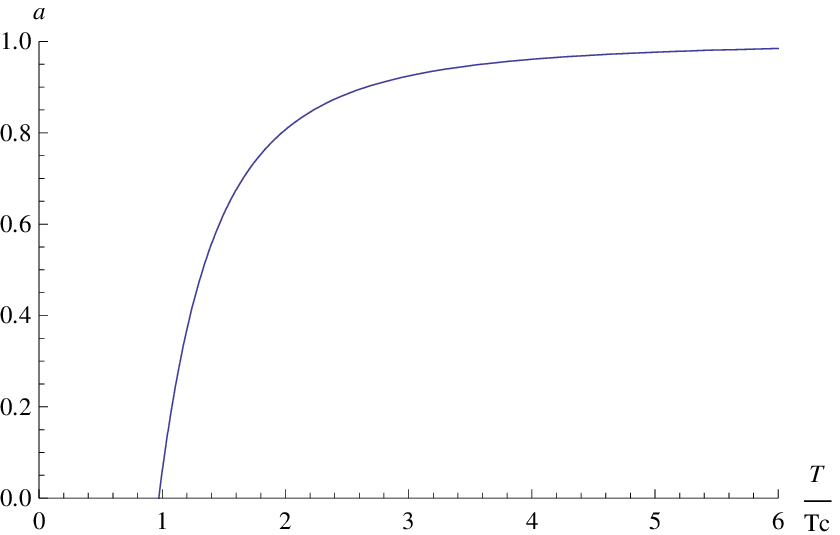}}
\caption{(a) Free Energy $F_1$,
(b)Internal energy $U_1$
both computed in Lattice QCD by Kakzmarek {\it et al}, (c) the fitting method,
and (d) the mixing energy parameter $a(T)$.}
\label{fig01} 
\end{figure}

\begin{figure}[t!]
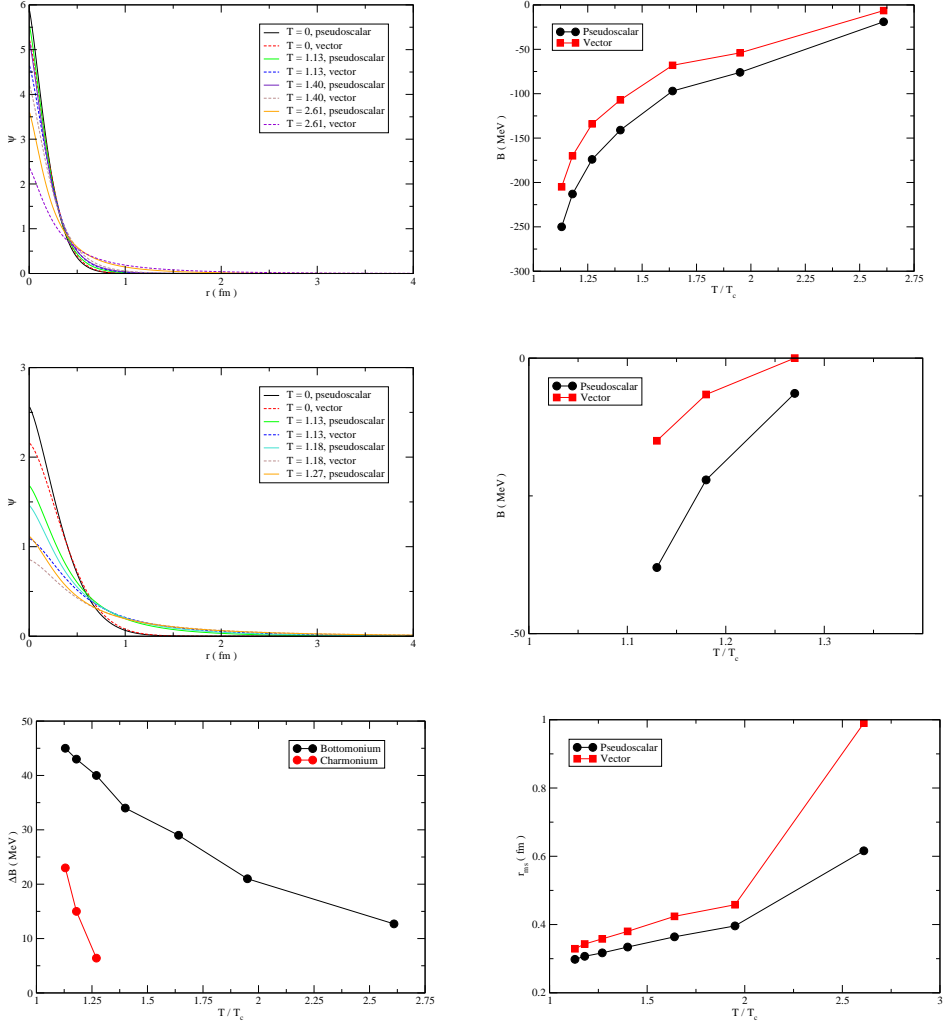

\includegraphics[width=0.43\columnwidth]{fig5_bottomwf.eps}
\hspace{.8cm}
\includegraphics[width=0.45\columnwidth]{fig6_binding_b.eps}
\\
\vspace{.2cm}
\\
\includegraphics[width=0.43\columnwidth]{fig7_charmwf.eps}
\hspace{.8cm}
\includegraphics[width=0.45\columnwidth]{fig8_binding_c.eps}
\\
\vspace{.2cm}
\\
\includegraphics[width=0.45\columnwidth]{fig9_diff_b.eps}
\hspace{.8cm}
\includegraphics[width=0.45\columnwidth]{fig10_rms_b.eps}
\caption{(a) bottomonium $\Upsilon$ and
$\eta_b$ wavefunctions,
(b) bottomonium $\Upsilon$ and
$\eta_b$ binding energies,
(c) charmonium $J\psi$ and $\eta_c$ wavefunctions and right,
(d) charmonium $J\psi$ and $\eta_c$ binding energies,
(e) bottonomium and charmonium hyperfine splitting $M_{vector}-M_{pseudoscalar}$
and (f) bottomonium $\upsilon$ and $\eta_b$ radius mean square.
}
\label{fig02} 
\end{figure}

The string confinement model
is dominant at moderate distances
while at short distances the attraction
is a Coulomb.
Olaf Kakzmarec {\it et al.  }
\cite{Doring:2007uh,Kaczmarek:2005ui}
fitted the lattice QCD static potential,
for spinless or infinitely heavy quarks,
\be
V(r) \to {- 4 \alpha \over 3 \, r} + \sigma \, r
\ee
with $\alpha= 0.212$  all in units of $\sqrt{\sigma}= 420 MeV$.
Including spin, at low distances the perturbative
One-Gluon-Exchange OGE produces a hyperfine
splitting.
As a first correction due to our heavy but not infinite quark masses,
we add the hyperfine $\mathbf{S}_1 \cdot \mathbf{S}_2 $ term that we fit
to the $J/\psi - \eta_c$  mass splitting and
to the $\upsilon - \eta_b$ mass splitting
(where we use the very recent Babar result
\cite{:2008vj}
of
$M_{\eta_b}=9388.9$ MeV/c$^2$),
\be
V(r) = {- 4 \alpha \left(1+ f
 \mathbf{S}_1 \cdot \mathbf{S}_2 \right)\over  3r}
+ V_0 + \sigma \, r
\ee
where $f_c=0.33$ and $f_b=0.12$.

At finite $T$, we use as thermodynamic
potentials, the free energy $F_1$ and the
internal energy $U_1$, computed in Lattice
QCD with the Polyakov loop
\cite{Doring:2007uh,Kaczmarek:2005ui}.
They are related to the static potential
$V(r)   = - f d r$
with
$F1 (r)= - f d r - S d T$
adequate for isothermic transformations
and with
$U1 (r)= - f d r + T d S$
adequate for adiabatic transformations.
The potentials are fitted
\cite{Bicudo:2008gs}
as described in Fig. \ref{fig01}.
We follow Cheuk-Yin Wong
\cite{Wong:2004zr},
who used the local gluon pressure
to combine the free energy $F_1$
and the Internal energy $U_1$
with the function $a(T)$
also depicted in Fig. \ref{fig01},
\be
V_T(r)=   { 3 \over   3 + a(T )  }         F1(r, T ) +{         a(T ) \over  3 + a(T )  }   U1(r, T ) \ .
\label{Va}
\ee

Solving the boundstate equation
\cite{Bicudo:2008gs}
for the potential $V_T(r)$ defined in eq. (\ref{Va}), we find the results show in Table \ref{table01} and in Fig. \ref{fig01}.

\begin{table}
\begin{center}
\begin{tabular}{ccccc}
\hline
$T/Tc$�����&� $B_0$�����&��� $B_{PS}$����&� $B_{V}$������&� $\Delta$ \\
\hline
1.13���&� -213����&� -250����& -205����&� 45  \\
1.18���&��� -178���&� -213����&� -170 ���&� 43  \\
1.27 ��& � -141 ��&�� -174 ���&� -134 ���&� 40  \\
1.40���&��� -112���&�� -141�����& -107����&� 34  \\
1.64���&���� -72����&��� -97����&���� -68���&� 29  \\
1.95���&���� -58����&��� -76�����&�� -54���&�� 21  \\
2.61���&���� -7.8����&��� -19�����&�� -6.4����& 12.7  \\
\hline
\end{tabular}
\end{center}
\caption{Table of the Bottonomium binding,
$B_0$ is the binding energy without hyperfine potential,
$B_{PS}$ is the binding energy for the Pseudoscalar,
$B_{V}$ is the binding energy for the Vector,
and $\Delta=B_{V}-B_{PS}$ is the hyperfine splitting.
}
\label{table01}
\end{table}

To conclude, we find that in $I=0$ quarkonium, the only existing boundstates above $T_c$
are the $L=0$ heavy quark groundstates with spin 1 and spin 0. The hyperfine potential splits the melting temperatures of the spin $S=J=1$ vector and the spin $S=J=0$ pseudoscalar.
 Bottomonium melts at higher temperatures, at $T= 2.7 \, T_c$ for $\Upsilon$
and $T=2.9 \, T_c$ for $\eta_c$, relevant for the LHC.
 Charmonium melts close to $T= T_c$, at $T= 1.2 \,T_c$ for $J/ \psi$
and $T= 1.3 \, T_c$ for $\eta_b$, rather cold for the LHC.

In the future, using modern quark model techniques, one might extend the present work to study light quark
chiral symmetry breaking and quark mass generation, compute the spectrum of any hadron and compute the interaction of any hadron-hadron,  at finite $T$.

\vspace{.1 cm}
{\bf Acknowledgement}
\\
We are very grateful to Olaf Kaczmarek for providing
his results for the static free energy and internal energy
in lattice QCD.

%
\end{document}